\documentclass[aps,prb,english,twocolumn,showpacs,preprintnumbers,amsmath,amssymb,floatfix]{revtex4-1}
\usepackage[T1]{fontenc}
\usepackage[latin9]{inputenc}
\usepackage{graphicx}
\usepackage{amssymb}
\usepackage{babel}
\makeatletter

\usepackage[version=3]{mhchem}

\usepackage{color}

\usepackage{ulem}

\makeatother
\usepackage{babel}
\makeatother

\begin{document}

\title{Core-level spectra from bilayer graphene}

\author{Bo E. Sernelius}

\email{bos@ifm.liu.se}
\affiliation{Division of Theory and Modeling, Department of Physics, Chemistry and Biology, Link\"{o}ping University, SE-581 83 Link\"{o}ping, Sweden}

\begin{abstract}
We derive core-level spectra for doped free-standing bilayer graphene. Numerical results are presented for all nine combinations of the doping concentrations $10^{12}  \rm{cm}^{-2}$, $10^{13}  \rm{cm}^{-2}$, and $10^{14}  \rm{cm}^{-2}$ in the two graphene sheets and we compare the results to the reference spectra for monolayer graphene. We furthermore discuss the spectrum of single-particle inter-band and intra-band excitations in the ${\omega}q$-plane, and show how the dispersion curves of the collective modes are modified in the bilayer system.
 \end{abstract}

\maketitle

\section{\label{introduction}Introduction}
In an earlier work\cite{Ser1} we derived core-level spectra for single free-standing graphene sheets. Our derivations were based on a model used by Langreth\,\cite{Lang} for the core-hole problem\,\cite{Lang1,Lang2,Gad,Citrin1,Citrin2,DonSun, Mahan} in the seventies. Our theoretical results very well reproduced experimental results, both for pristine and doped graphene. 

In the excitation process the photoelectron leaves the system and a core hole is left behind. The shape of the XPS (x-ray photoelectron spectroscopy) spectrum depends on how fast the process is. One may use the adiabatic approximation if the process is very slow. In this approximation one assumes that the electrons in the system have time to relax around the core hole during the process. When we derive the XPS line shape we go to the other extreme and assume that the excitation process is very fast; we use the sudden approximation. In this approximation the core-hole potential is turned on instantaneously. The electrons have not time, during the process, to settle down and reach equilibrium in the potential caused by the sudden appearance of  the core hole. This results in shake-up effects in the form of single particle (electron-hole pair) excitations and collective (plasmon) excitations. The electrons contributing to the shake-up effects are the electrons in the valence and conduction bands. We use the conical approximation for these bands in which both bands have conical shapes for all momenta. We make the assumption that the core hole does not recoil in the shake-up process and that there are no excitations within the core. We approximate the core-hole potential with a pure Coulomb potential. Furthermore, the finite lifetime of the core hole causes a Lorentzian broadening of all peaks and experimental uncertainties give a Gaussian broadening.

In many experiments one finds samples with more than one graphene sheet and the doping levels are often different in the sheets. This motivates the present work where we investigate how the core-level spectra in a bilayer system  differ from that in a monolayer system. The sudden creation of the core hole in one layer is expected to cause shake-up processes not just in that layer but also in the other layer. To be more precise, the collective excitations in the two layers are coupled and the spectra are affected by single-particle excitations in the two layers and by collective excitations of the bilayer system.

The material is organized in the following way: In Sec. \ref{the} we summarize the formulas for the core-level spectra derived for a 2D system in Ref.\,[\onlinecite{Ser1}] and show how they are modified for a bilayer system. We furthermore discuss the spectrum of single-particle inter-band and intra-band excitations in the  ${\omega}q$-plane, and show how the dispersion curves of the collective modes are modified in the bilayer system. The numerical results for the core-level spectra are presented in Sec. \ref{res}. Finally, we end with a brief summary and conclusion section, Sec. \ref{sum}.

\section{\label{the}Theory}

We found in Ref.\,[\onlinecite{Ser1}] that the XPS spectrum for a core-level in a single graphene sheet can be written as
\begin{equation}
\begin{array}{l}
S(W) = \frac{1}{\pi }\int\limits_0^\infty  {{e^{\left( { - \frac{{LW}}{2}T} \right)}}{e^{ - a(T)}}} \\
\quad \quad \quad  \times \cos \left[ {\left( {W - D} \right)T - b\left( T \right)} \right]dT,
\end{array}
\label{equ1}
\end{equation}
where
\begin{equation}
\begin{array}{*{20}{l}}
{a(T) = \frac{{{e^2}{k_F}}}{{\pi {E_F}}}\int\limits_0^\infty  {\int\limits_0^\infty  {\frac{1}{{{W^2}}}{\rm{Im}}\left[ {\frac{{ - 1}}{{\varepsilon \left( {Q,W} \right)}}} \right]} } }\\
{\quad \quad \quad  \times \left[ {1 - \cos \left( {WT} \right)} \right]dQdW,}
\end{array}
\label{equ2}
\end{equation}
\begin{equation}
\begin{array}{l}
b(T) =  - \frac{{{e^2}{k_F}}}{{\pi {E_F}}}\int\limits_0^\infty  {\int\limits_0^\infty  {\frac{1}{{{W^2}}}{\rm{Im}}\left[ {\frac{{ - 1}}{{\varepsilon \left( {Q,W} \right)}}} \right]} } \\
\quad \quad \quad  \times \sin \left( {WT} \right)dQdW,
\end{array}
\label{equ3}
\end{equation}
and
\begin{equation}
D = \frac{{{e^2}{k_F}}}{{\pi {E_F}}}\int\limits_0^\infty  {\int\limits_0^\infty  {\frac{1}{W}{\rm{Im}}\left[ {\frac{{ - 1}}{{\varepsilon \left( {Q,W} \right)}}} \right]dQ} dW}.
\label{equ4}
\end{equation}

All variables have been scaled according to
\begin{equation}
\begin{array}{l}
Q = q/2{k_F},
W = \hbar \omega /2{E_F},\\
T = t2{E_F}/\hbar ,
D = d/2{E_F},\\
LW = lw/2{E_F},
GW = gw/2{E_F},\\
{W_0} = \hbar {\omega _0}/2{E_F},
\end{array}
\label{equ5}
\end{equation}
and are now dimensionless. The function $\varepsilon \left( {{\bf{q}},\omega } \right)$ is the dielectric function of the graphene sheet. It is a function of the 2D (two-dimensional) momentum ${\bf{q}}$ and frequency $\omega$. The analytical expression for this function in graphene is given in Eq. (46) of Ref.\,[\onlinecite{Ser1}]. The quantities ${E_F}$ and ${k_F}$ are the Fermi energy and Fermi wave-number, respectively, for the doping carriers. The quantity $d$ is the energy shift of the adiabatic peak and $t$ is the time variable. We have taken the finite life-time of the core hole into account by introducing the first factor of the integrand of Eq.(\ref{equ1}), where ${lw}$ is the FWHM (Full Width at Half Maximum) of the Lorentz broadened peak. The gaussian instrumental broadening, with a FWHM of ${gw}$, is also taken into account using the same trick as in our earlier work\cite{Ser1}. In the numerical results presented below we use the same parameter values as in Fig.\,7 of Ref.\,[\onlinecite{Ser1}], i.e., ${lw}$ and ${gw}$ have been given the values .12 eV and .305 eV, respectively.

 We have assumed that the core-hole potential can be represented by a pure Coulomb potential. The results are valid for a general 2D system. The particular system enters the problem through ${\mathop{\rm Im}\nolimits} \left[ {\varepsilon {{\left( {Q,W} \right)}^{ - 1}}} \right]$.

The system we treat here consists of two graphene sheets, sheet number 1 and sheet number 2, separated by a distance $\delta$ with the value\cite{Moh} 3.35{\rm{{\AA}}}. We let the core-level be in sheet number 1. The core hole will be dynamically screened by the carriers in both sheet number 1 and 2. The resulting potential when both sheets have the same doping concentration was given in Ref.[\onlinecite{Ser2}]. Here we give a more general result valid for independent doping levels in the two sheets. A Coulomb potential ${{v^{2D}}\left( q \right)}$ in layer $i$ results in the dynamically screened potential ${{\tilde v}_{ij}}\left( {{\bf{q}},\omega } \right)$ in layer $j$, where
\begin{equation}
\begin{array}{l}
{{\tilde v}_{11}}\left( {{\bf{q}},\omega } \right) = \frac{{{v^{2D}}\left( q \right)\left\{ {1 + {\alpha _2}\left( {{\bf{q}},\omega } \right)\left[ {1 - \exp \left( { - 2q\delta } \right)} \right]} \right\}}}{{1 + {\alpha _1}\left( {{\bf{q}},\omega } \right) + {\alpha _2}\left( {{\bf{q}},\omega } \right) + {\alpha _1}\left( {{\bf{q}},\omega } \right){\alpha _2}\left( {{\bf{q}},\omega } \right)\left[ {1 - \exp \left( { - 2q\delta } \right)} \right]}},\\
{{\tilde v}_{22}}\left( {{\bf{q}},\omega } \right) = \frac{{{v^{2D}}\left( q \right)\left\{ {1 + {\alpha _1}\left( {{\bf{q}},\omega } \right)\left[ {1 - \exp \left( { - 2q\delta } \right)} \right]} \right\}}}{{1 + {\alpha _1}\left( {{\bf{q}},\omega } \right) + {\alpha _2}\left( {{\bf{q}},\omega } \right) + {\alpha _1}\left( {{\bf{q}},\omega } \right){\alpha _2}\left( {{\bf{q}},\omega } \right)\left[ {1 - \exp \left( { - 2q\delta } \right)} \right]}},\\
{{\tilde v}_{12}}\left( {{\bf{q}},\omega } \right) = \frac{{{v^{2D}}\left( q \right)\exp \left( { - q\delta } \right)}}{{1 + {\alpha _1}\left( {{\bf{q}},\omega } \right) + {\alpha _2}\left( {{\bf{q}},\omega } \right) + {\alpha _1}\left( {{\bf{q}},\omega } \right){\alpha _2}\left( {{\bf{q}},\omega } \right)\left[ {1 - \exp \left( { - 2q\delta } \right)} \right]}}.
\end{array}
\label{equ6}
\end{equation}
Since we have a background screening constant $\kappa$ all $v$- and $\alpha$- functions should be divided by this constant and ${\varepsilon _i}\left( {Q,W} \right) = \kappa \left[ {1 + {\alpha _i}\left( {Q,W} \right)/\kappa } \right]$.  All expressions in Eqs.\,(\ref{equ2})-(\ref{equ4})valid for a single graphene sheet are now valid for a bilayer after the replacement:
\begin{equation}
\begin{array}{l}
{\mathop{\rm Im}\nolimits} \left[ {\frac{{ - 1}}{{\varepsilon \left( {Q,W} \right)}}} \right] \to \\
 - {\mathop{\rm Im}\nolimits} \frac{{{\varepsilon _2}\left( {Q,W} \right) - \left[ {{\varepsilon _2}\left( {Q,W} \right) - \kappa } \right]\exp \left( { - 2Q\Delta } \right)}}{{{\varepsilon _1}\left( {Q,W} \right){\varepsilon _2}\left( {Q,W} \right) - \left[ {{\varepsilon _1}\left( {Q,W} \right) - \kappa } \right]\left[ {{\varepsilon _2}\left( {Q,W} \right) - \kappa } \right]\exp \left( { - 2Q\Delta } \right)}},
\end{array}
\label{equ7}
\end{equation}
where $\Delta  = \delta 2{k_F}$.

\begin{figure}
\includegraphics[width=8.5cm]{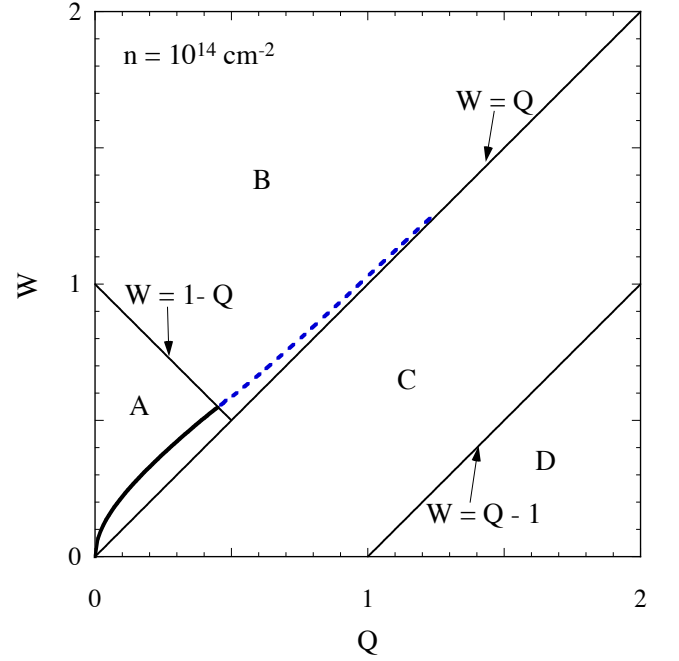}
\caption{(Color online) Excitation domains in a single graphene sheet with doping density $10^{14}  \rm{cm}^{-2}$. In a  pristine graphene sheet there are inter-band single-particle excitations in the whole A- and B-regions of the WQ-plane. There are no collective excitations and no intra-band single-particle excitations; there are no excitations below the $W = Q$ line. When the sheet is doped some inter-band excitations are prevented since they are blocked by the doping carriers (Pauli's exclusion principle); there are no longer any inter-band excitations in region A , i.e., below the $W=1-Q$ line. There are now intra-band single-particle excitations; these are limited to region C. Furthermore there are collective excitations, plasmons, above the $W = Q$ line. There is a distinct plasmon curve in region A since this region is free from single-particle excitations; when the plasmon curve inters region B it broadens and either eventually merges with the intra-band single-particle continuum or stops at some point. }
\label{figu1}
\end{figure}

We now have all we need for the calculation of the core-hole spectra but before we present the numerical results let us discuss the possible excitations of the bilayer system. We limit the discussion to electronic excitations and first begin with a monolayer system. 

In a single free-standing pristine graphene sheet there are only single-particle excitations, no collective excitations. Due to the conical shapes of the conduction- and valence-bands these all fall above the $W = Q$ line in Fig.\,\ref{figu1}. The inter-band continuum covers regions $A$ and $B$ and there are no excitations in regions $C$ and $D$. When the sheet is doped the inter-band continuum is depleted and there are also new excitation types, viz., single-particle intra-band excitations and collective excitations. If the doping is $n$-type electrons are filling up the states at the bottom of the conduction band. This excludes some of the inter-band transitions since Pauli's exclusion principle prevents electrons excited from the valence band to end up in these occupied conduction-band states. This leads to a depletion of the continuum above the $W = Q$ line. Actually, the continuum vanishes completely in region $A$. There are new single-particle excitations where the doping electrons in the conduction band are excited to states further up in the band. These excitations form a continuum covering region $C$. There are also collective excitations above the $W = Q$ line. These are plasmons with the dispersion curve starting out for small momenta as $W \sim \sqrt Q $, characteristic of a 2D system. Since there are no single particle continuum in region $A$ there is a distinct plasmon curve in this region. For larger momenta this curve enters region $B$ where the curve broadens, due to the single-particle continuum, and eventually stops. This broadening is represented in the figure by making the curve dashed.  The curve in the figure is for a doping concentration of $10^{14}  \rm{cm}^{-2}$. Since the band-structure is symmetric in energy the results are equally valid for $p$-type doping. In that case electrons are absent from the states at the top of the valence band. This reduces the amount of inter-band excitation since the excitations from these states into the conduction band no longer can take place. New intra-band excitations occur where electron further down in the valence band can be excited into these now empty states.

\begin{figure}
\includegraphics[width=8.5cm]{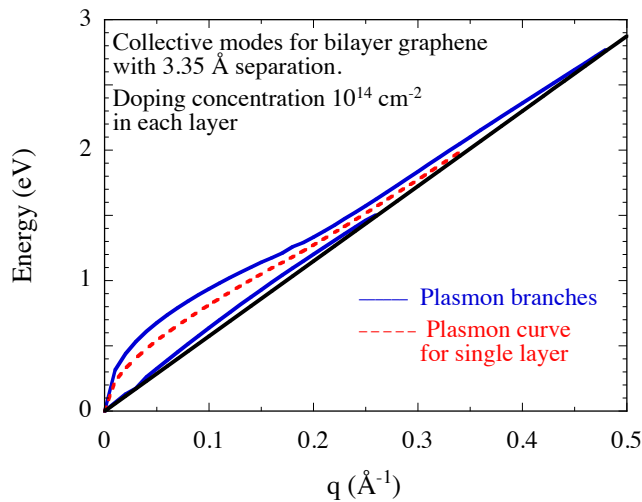}
\caption{(Color online) Dispersion curves for the collective modes of bilayer graphene where both layers have a doping concentration of  $10^{14}  \rm{cm}^{-2}$. The solid blue curves are the two plasmon branches when the layers have a separation of 3.35 {\rm{{\AA}}}. The red short-dashed curve is the plasmon curve for a single layer. When the separation between the two layers increases towards infinity the two plasmon branches come together and  become degenerate and coincide with this curve in the limit. The black solid straight line is the upper boundary of the intra-band single-particle continuum.}
\label{figu2}
\end{figure}

\begin{figure}
\includegraphics[width=8.5cm]{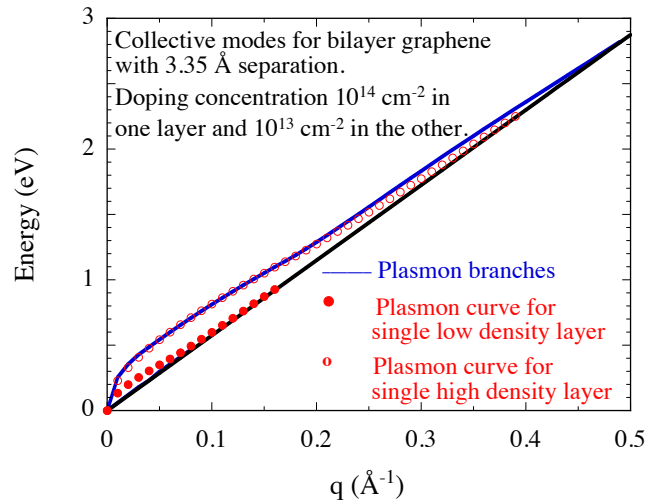}
\caption{(Color online) The same as Fig.\,\ref{figu2} but now for a bilayer where one layer has the doping concentration $10^{14}  \rm{cm}^{-2}$ and the other $10^{13}  \rm{cm}^{-2}$. The solid circles show the dispersion curve for plasmons in a single graphene layer with doping concentration $10^{13}  \rm{cm}^{-2}$ while the open circles show the corresponding for the doping concentration $10^{14}  \rm{cm}^{-2}$.  }
\label{figu3}
\end{figure}

In a bilayer system there are single-particle excitations in both sheets and if one or both sheets are doped there are also collective excitations. If the sheets are well separated the excitations in the two sheets are independent of each other. When the separation decreases the single particle excitation continua in the two sheets are unchanged. However, the coupling constants for processes where the excitations are a part change. This is, e.g., important for core-level spectroscopy, addressed in the present work. The effect of coupling is even more dramatic for the collective modes.   Here, apart from the change in coupling constants the dispersion curves for the modes change. If the doping concentration is the same in the two sheets the dispersion curves for the plasmons in the two sheets are degenerate for large separation. When the separation decreases the two plasmon modes split up and the collective modes for the bilayer system will have two branches. 

This is illustrated in Fig.\,\ref{figu2} where we show the dispersion curves for the collective modes for a graphene bilayer with two graphene sheets separated by 3.35 {\rm{{\AA}}}, both with a doping concentration of ${10^{14}}{\rm{c}}{{\rm{m}}^{ - 2}}$. When the two layers are far apart the plasmon dispersion curves (red short dashed curve) for the two layers are degenerate. When the layers are brought together the curves split up and form two plasmon branches (blue solid curves). The curves merge with the single-particle intra-band continuum for large momenta or stop before that; they may stop before merging since they are already inside a continuum, the inter-band continuum. The black solid straight line represents the upper boundary of the  intra-band continuum.
If the two layers have different amount of doping, for large separations the two plasmon dispersion curves are not degenerate. When the sheets are brought together the two branches of the collective modes of the bilayer system start out as the two single-sheet plasmon-modes but the closer the sheets come the stronger the coupling and the more the two dispersion curves are modified. 

This is illustrated in Fig.\,\ref{figu3} for a graphene bilayer with the doping concentration $10^{14}  \rm{cm}^{-2}$ in one of the sheets and  $10^{13}  \rm{cm}^{-2}$ in the other. Now, the plasmon curves for the individual layers are no longer degenerate. When the sheets are brought together the sheets couple to each other and two new dispersion curves bracket the two individual plasmon curves. The upper branch follows very closely to the high-density plasmon curve and the lower branch is squeezed between the low-density plasmon curve and the continuum. It stays above the continuum for small momenta but merges fairly quickly with the continuum with increasing momentum.  

In Figs. \ref{figu2} and  \ref{figu3} we have not discussed damping of the plasmon branches. It becomes more complicated when the doping concentration is different in the two sheets. The regions $A$ of Fig.\,\ref{figu1} for the two sheets would cover different regions when transferred to Fig.\,\ref{figu3}. Distinct (undoped) plasmon curves only appear in the smaller $A$-region defined by the low density sheet. For our calculations here we do not have to bother with the damping and finding the dispersion curves. This is all included in the formalism.

\section{\label{res}Results}

\begin{figure}
\includegraphics[width=8.5cm]{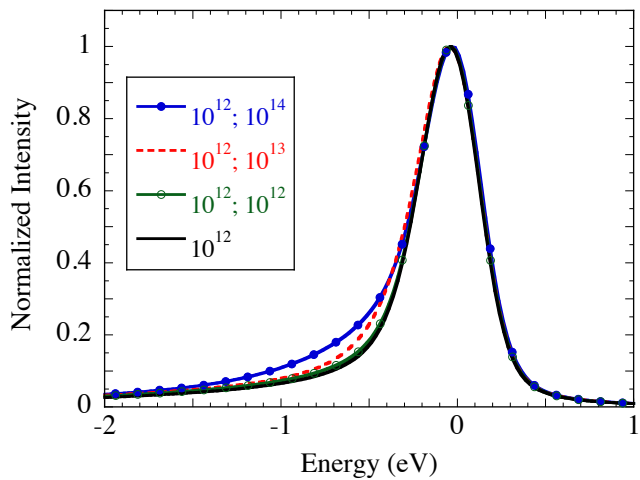}
\caption{(Color online) The shape of the C 1s core-level spectra of bilayer graphene. To get a better view of the peak shapes we have removed the over all shift, $d$, from all spectra.  The black solid curve, the reference curve,  is the result from a single free-standing graphene sheet with doping concentration $10^{12} {\rm cm}^{-2}$; the green solid curve dressed with open circles is the result when another sheet with the same concentration is put next to it; the red dashed curve is the result when the doping concentration of the second sheet is increased to  $10^{13} {\rm cm}^{-2}$; the blue curve dressed with solid circles is  the result when the doping concentration of the second sheet is increased to  $10^{14} {\rm cm}^{-2}$.}
\label{figu4}
\end{figure}

\begin{figure}
\includegraphics[width=8.5cm]{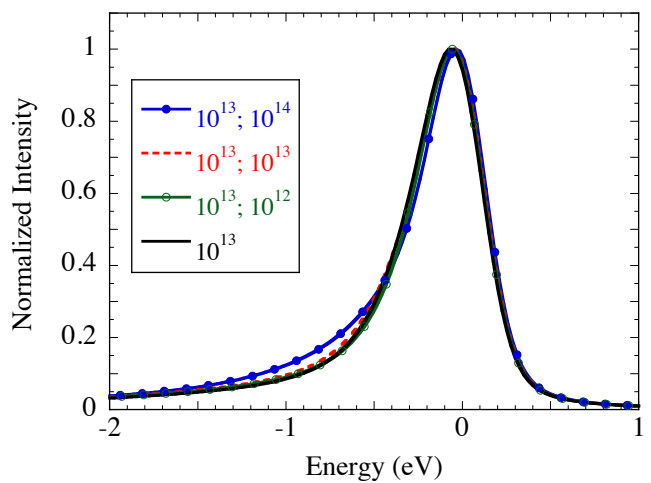}
\caption{(Color online) The shape of the C 1s core-level spectra of bilayer graphene. To get a better view of the peak shapes we have removed the over all shift, $d$, from all spectra.  The black solid curve, the reference curve,  is the result from a single free-standing graphene sheet with doping concentration $10^{13} {\rm cm}^{-2}$; the green solid curve dressed with open circles is the result when another sheet with doping concentration $10^{12} {\rm cm}^{-2}$ is put next to it; the red dashed curve is the result when the doping concentration of the second sheet is increased to  $10^{13} {\rm cm}^{-2}$; the blue curve dressed with solid circles is  the result when the doping concentration of the second sheet is increased to  $10^{14} {\rm cm}^{-2}$.}
\label{figu5}
\end{figure}

In a core-level spectrum from bilayer graphene there will be two overlapping contributions, one from when the core-level is in one of the sheets and one from when the core-level is in the other. These two contributions are shifted relative each other by the Fermi energy difference in the two sheets and by the difference in interaction shifts. We have here concentrated on the peak shapes only and not on the position of the peaks. 
In each of the figures we present we have chosen a fix doping concentration in the sheet containing the core-level. We show the peak for a single free-standing sheet with this doping concentration as a reference peak. Then we give the result when a second sheet, with the same or different doping concentration, is put close to the sheet containing the core-level. We have chosen the separation to be  3.35{\rm{{\AA}}} \cite{Moh} in all examples. A larger value would give a smaller effect on the peaks and smaller value a larger effect. The first number in the figure legends denotes the dopant concentration in units of $\rm{cm}^{-2}$ for the sheet containing the core hole and the second number is for the other sheet.

In Fig.\,\ref{figu4} the sheet containing the core hole has the doping concentration $10^{12} {\rm cm}^{-2}$. We find that a second sheet with the 
the same doping concentration has a very small effect on the peak shape. Increasing the concentration in the second sheet to $10^{13} {\rm cm}^{-2}$ gives a noticeable enhancement on the low energy side of the peak. Further increase of the doping concentration to $10^{14} {\rm cm}^{-2}$ gives an additional enhancement now shifted towards lower energies. 

Fig.\,\ref{figu5} shows the corresponding results when the sheet containing the core hole has the doping concentration $10^{13} {\rm cm}^{-2}$. We find that a second sheet with the doping concentration $10^{12} {\rm cm}^{-2}$ or $10^{13} {\rm cm}^{-2}$ has a very small effect on the peak shape. Increasing the concentration in the second sheet to $10^{14} {\rm cm}^{-2}$ gives a noticeable enhancement on the low energy side of the peak. 

\begin{figure}
\includegraphics[width=8.5cm]{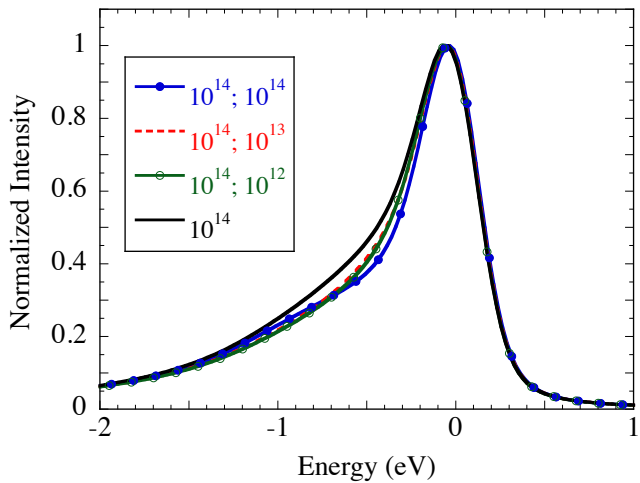}
\caption{(Color online) The shape of the C 1s core-level spectra of bilayer graphene. To get a better view of the peak shapes we have removed the over all shift, $d$, from all spectra.  The black solid curve, the reference curve,  is the result from a single free-standing graphene sheet with doping concentration $10^{14} {\rm cm}^{-2}$; the green solid curve dressed with open circles is the result when another sheet with doping concentration $10^{12} {\rm cm}^{-2}$ is put next to it; the red dashed curve is the result when the doping concentration of the second sheet is increased to  $10^{13} {\rm cm}^{-2}$; the blue curve dressed with solid circles is  the result when the doping concentration of the second sheet is increased to  $10^{14} {\rm cm}^{-2}$. }
\label{figu6}
\end{figure}

Finally, in our last example, Fig.\,\ref{figu6}, we continue to increase the doping concentration in the sheet with the core hole, now to $10^{14} {\rm cm}^{-2}$. The reference peak in this case has a well developed shoulder. When bringing in a second sheet with doping doping concentration $10^{12} {\rm cm}^{-2}$ actually reduces this shoulder. When the concentration in the second sheet is increased to $10^{13} {\rm cm}^{-2}$ very little happens. Finally, when the concentration in the second sheet is increased to $10^{14} {\rm cm}^{-2}$ the shoulder is reduced further in a region of small negative energies   but comes back up further down in the tail region. This example demonstrates a different and unexpected behavior and shows that it is difficult to predict the outcome. Simulations are necessary.

\section{\label{sum}Summary and Conclusions}

We have derived the formula for the core-level spectra in bilayer graphene and presented numerical spectra for a combination of doping concentrations in the two graphene sheets. In a real experiment with doping concentrations ${n_i}$ and ${n_j}$ in the two sheets the spectrum consists of the superposition of the two spectra  ${n_i}$;${n_j}$ and ${n_j}$;${n_i}$ with the notation used in the legends of  Figs.\,\ref{figu4}-\ref{figu6}. When the two spectra are superimposed they should be shifted in energy relative each other by the Fermi energy difference in the two sheets and by the difference in the interaction shifts. In the formalism we use one obtains the interaction shifts, $d$, but we have refrained from giving them here because they are sensitive to the band-structure approximation. In the real band structure the valence- and conduction-bands deviate from conical for energies a couple of electron volts away from the Dirac point where the bands touch. Thus, the true location of the peaks is sensitive to deviations of the band shapes from conical. This is expected to influence the results more the higher the doping concentration. 

When analyzing the results shown in Figs.\,\ref{figu4}-\ref{figu6} we found that for the two first figures the tail to the left in the spectrum increased when a second sheet was introduced. In Fig.\,\ref{figu6} the opposite was found. One should keep in mind that the peaks are normalized in such a way that the peak maximum is put equal to unity, This means that an increase in small-energy shake-up processes will lower the values in the shoulder- and tail-regions.

All the results presented here are for free-standing bilayers. When a bilayer is on a substrate the results may be affected both by screening from the substrate and by possible additional shake-up processes taken place in the substrate.



\begin{thebibliography}{10}


\bibitem{Ser1} Bo E. Sernelius, Phys. Rev. B {\bf 91}, 045402(2015).
\bibitem{Lang} David C. Langreth, Phys. Rev. B {\bf 1}, 471 (1970).

\bibitem{Lang1}Jhy-Jiun Chang and David C. Langreth, Phys. Rev. B {\bf 5}, 3512 (1972).
\bibitem{Lang2}Jhy-Jiun Chang and David C. Langreth, Phys. Rev. B  {\bf 8}, 4638 (1973).
\bibitem{Gad}J. W. Gadzuk and M. \u{S}unji\'{c}, Phys. Rev. B  {\bf12}, 524 (1975).

\bibitem{Citrin1}P. H. Citrin, G. K Wertheim, and Y. Baer, Phys. Rev. Lett.  {\bf 35}, 885 (1975).
\bibitem{Citrin2}P. H. Citrin, G. K Wertheim, and Y. Baer, Phys. Rev. Lett.  {\bf 41}, 1425 (1978).
\bibitem{DonSun} Doniach and M \u{S}unji\'{c}, J. Phys. C: Solid State Phys. {\bf 3}, 285 (1970).

\bibitem{Mahan}G. D. Mahan, Phys. Rev. B  {\bf 11}, 4814 (1975).

\bibitem{Moh}M. S. Alam, J. Lin, and M. Saito, Jap. J. of Appl. Phys. {\bf 50}, 080213 (2011). 
\bibitem{Ser2} Bo E. Sernelius, J. Phys.: Condens. Matter {\bf 27}, 214017 (2015).

\end{thebibliography}
\end{document}